# Revisiting Circular-Based Random Node Simulation

Mouhamed Abdulla[*], Yousef R. Shayan[*] and Junho Baek[†]

[*]Department of Electrical and Computer Engineering, Concordia University, Montréal, Québec, Canada
[†]Department of Electronics Engineering, Dongguk University, Seoul, Korea
E-mail: {m_abdull, yshayan}@ece.concordia.ca[*] and {iclicker}@dongguk.edu[†]

*Abstract*—In literature, a stochastic model for spreading nodes in a cellular cell is available. Despite its existence, the current method does not offer any versatility in dealing with sectored layers. Of course, this needed adaptability could be created synthetically through heuristic means. However, due to selective sampling, such practice dissolves the true randomness sought. Hence, in this paper, a universal exact scattering model is derived. Also, as an alternative to exhaustive simulation, a generic close-form path-loss predictor between a node and a BS is obtained. Further, using these results, an algorithm based on the superposition principle is proposed. This will ensure greater emulation flexibility, and attain a heterogeneous spatial density.

## I. INTRODUCTION

The importance of cellular technology is still, and even more significant as move toward 4G systems with WiMAX, LTE, and MBWA [1]. Therefore, time and cost efficient analysis of such systems based on simulation is equally vital.

In this paper, the emphasis will be on random emulation of terminals spatial position. And, such study is conducted because the emplacement of nodes will impact factors such as: network capacity, coverage area, connectivity, resource consumption, interference, etc. Hence, it is a common practice by researchers and design engineers to spread random nodes in a circular cell [2] [3], representing an ideal EM radiation profile of a Base Station (BS). Also, distribution in an annulus is often conducted to study edge related aspects of a contour, among others [3] [4]. However, to the best of our knowledge, no comprehensive *exact stochastic* method exists for dropping nodes in different mutations of a circular cell, without relying on biased sampling. Thus, in section II, a universal model to achieve this will be derived such that randomness is preserved. Then, in Section III, a path-loss predictor for this new model will be obtained as a generalization to that shown in [5]. Finally in Section IV, using these results, an algorithm called CSA, with greater node spreading flexibility, will be proposed.

## II. A UNIVERSAL RANDOM SCATTERING MODEL

Consider Fig. 1, where $0 \leq L_1 < L_2$ are radii in units of length, and $0 \leq \alpha_1 < \alpha_2 \leq 2\pi$ are angular limits, all in $\mathbb{R}_+^4$. If we assume uniform spreading in the area of interest "$D$", then the joint Probability Density Function (PDF) becomes the reciprocal of this surface. Next, for ease of analysis, after transforming this PDF to polar coordinates, we find:

$$f_{R\theta}(r,\theta) = \left[f_{XY}(x,y)\right]_{\substack{x=r\cos(\theta)\\y=r\sin(\theta)}} |J(r,\theta)| = \frac{2r}{(L_2^2 - L_1^2)(\alpha_2 - \alpha_1)} \quad (1)$$

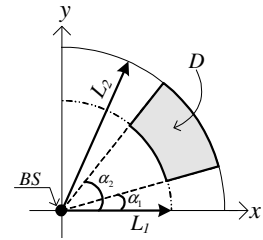

Fig. 1. A circular sector layer.

where $J(r,\theta)$ is a 2D Jacobian matrix, such that $0 \leq L_1 \leq r \leq L_2$ and $0 \leq \alpha_1 \leq \theta \leq \alpha_2 \leq 2\pi$ for $(r,\theta) \in D$. Now, with (1), the Cumulative Distribution Function (CDF) along "$r$" becomes:

$$F_R(r) = \int_{t=-\infty}^{r} \int_{\theta=\alpha_1}^{\alpha_2} f_{R\theta}(t,\theta) d\theta dt = \begin{cases} 0 & r \leq L_1 \\ \dfrac{(r^2 - L_1^2)}{(L_2^2 - L_1^2)} & L_1 \leq r \leq L_2 \\ 1 & r \geq L_2 \end{cases} \quad (2)$$

Further, using the inverse transformation algorithm, the CDF in (2) is made equal to some random sample "$u_1$" generated from a uniform density between zero and one, i.e. $U(0,1)$. Then, the inverse CDF, which is $\in \mathbb{R}_+$, becomes: $r = F_R^{-1}(u) = \sqrt{u_1(L_2^2 - L_1^2) + L_1^2}$. As for the angle, it is produced by: $\theta = u_2(\alpha_2 - \alpha_1) + \alpha_1$, where "$u_2$" is also obtained from $U(0,1)$, though uncorrelated from "$u_1$". Fig. 2 demonstrates the versatility of this model.

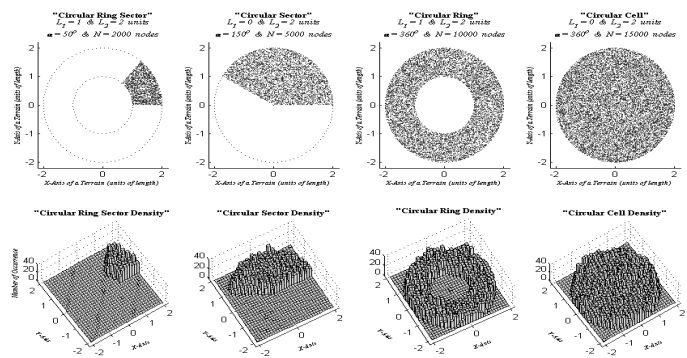

Fig. 2. Exact stochastic scattering of terminals.

## III. A GENERIC PATH-LOSS PREDICTOR

Being able to predict the density of a Path-Loss (PL) is always essential in wireless communications. Typically, the way to do this relies on intense Monte Carlo simulation.

Although here, a generic close-form PL predictor for the model derived in Section II will be shown analytically.

In general, PL with distance dependency is given by $L_P = \alpha + \beta \log_{10}(r/r_0) + \Psi$, where "$\alpha$" and "$\beta$" rely on the specific mobile system used, "$r_0$" is the close-in distance, "$r$" is the separation between a node and a BS, and "$\Psi$" is the contribution due to shadowing with a zero-mean Gaussian distribution and variance "$\sigma_\Psi$". Thus, from the inside of (2) we get the PDF for the radius; then, since "$\alpha, \beta, r_0$" are deterministic values and shadowing is independent from random variable "$r$", hence convolution is used to get:

$$f_{L_P}(l, \vec{\Lambda}) = \int_{\tau=l-w_2}^{l-w_1} N(0, \sigma_\Psi^2)\left[\ln(10) r f_R(r)/\beta\right]_{r=r_0 10^{(l-\tau-\alpha)/\beta}} d\tau \quad (3)$$

where $w_i = \alpha + \beta \log_{10}(L_i/r_0)$, $i=1,2$ and $\vec{\Lambda} = [L_1; L_2; d_0; \alpha; \beta; \sigma_\Psi]$ for $l \in \mathbb{R}$. After solving (3), we find a generic PDF for the PL:

$$f_{L_P}(l, \vec{\Lambda}) = \frac{2 r_0^2 \ln(10)}{\beta(L_2^2 - L_1^2)} 10^{2(\ln(10)\sigma_\Psi^2 + \beta(l-\alpha))/\beta^2} Q\left(\frac{\Omega - \beta \log_{10}(r)}{\sigma_\Psi}\right)\Bigg|_{r=L_1}^{r=L_2} \quad (4)$$

where $\Omega = l - \alpha + \beta \log_{10}(r_0) + 2\ln(10)\sigma_\Psi^2/\beta$, and "$Q$" is a variation of the complementary error function: $Q(x) = \text{erfc}(x/\sqrt{2})/2$. Fig. 3 shows the agreement between simulation and theory using values from the new IEEE 802.20.

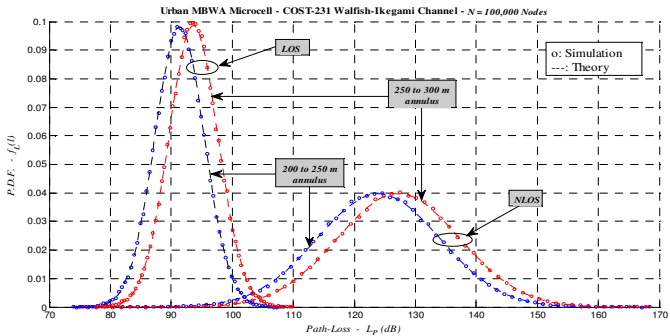

Fig. 3. PL simulation using MBWA channel environments.

## IV. CELLULAR-BASED SUPERPOSITION ALGORITHM – CSA

The model derived and analyzed above, gives the possibility to spread nodes unbiasedly in a specific sector. At this level, we may harness these findings and create an algorithm that gives the freedom to a cellular analyst or designer to selectively spread nodes in locations desired. This is done, because diverse users clustering schemes are caused by manmade and natural terrain features. The approach is based on splitting a circular cell into "$L$" *onion*-like layers; and then partition each layer into "$S_i$" sectors, where $i=1,2,..,L$. Totally, there are "$\gamma$" sectors, and in each sector "$n_m$" nodes are spread where $m=1,2,..,\gamma$. Once the simulation dropping is complete, after several iterations, the superposition principle can be applied to get the overall spatial distribution of the cell, hence the name CSA. This method is a simple emulation tool that proves useful, and achieves non-homogeneity, when knowledge about a cell site is known or hypothesized.

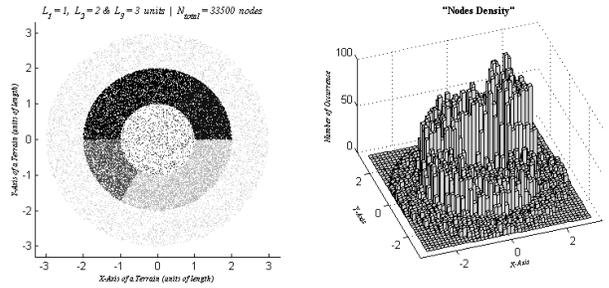

Fig. 4. Heterogeneous spatial distribution using the CSA algorithm.

TABLE I - CSA ALGORITHM

| | |
|---|---|
| 00 $t_1 = 0; t_2 = 0$ | $I_{1\times(L+1)} = [L, S_1, S_2, \cdots, S_L] \in \mathbb{N}_*^{L+1}$ |
| 01 **for** $i = 1, 2, \cdots, L$ | $\gamma = \sum_{l=1}^{L} S_l \in \mathbb{N}_*$ **Inputs** |
| 02 $\quad r_L = \Lambda_{(1,i)}; r_H = \Lambda_{(2,i)}$ | $\Lambda_{3\times(L+\gamma)} = \begin{bmatrix} R & \Theta \\ \bigcirc & \aleph \end{bmatrix}$ |
| 03 $\quad$ **for** $j = 1, 2, \cdots, S_i$ | |
| 04 $\qquad \theta_L = \Lambda_{(1,j+t_1+L)}; \theta_H = \Lambda_{(2,j+t_1+L)}$ | $R_{2\times L} = \begin{bmatrix} r_{11} & r_{12} & \cdots & r_{1L} \\ r_{21} & r_{22} & \cdots & r_{2L} \end{bmatrix} \in \mathbb{R}_+^{2\times L}$ |
| 05 $\qquad n = \Lambda_{(3, j+t_1+L)}$ | |
| 06 $\qquad$ **for** $k = 1, 2, \cdots, n$ | $\Theta_{2\times\gamma} = [\Theta^{(1)}_{2\times S_1} \Theta^{(2)}_{2\times S_2} \cdots \Theta^{(L)}_{2\times S_L}]$ |
| 07 $\qquad\quad u_1 \sim U(0,1); u_2 \sim U(0,1)$ | |
| 08 $\qquad\quad r_0 = \sqrt{u_1(r_H^2 - r_L^2) + r_L^2}$ | $\Theta^{(i)}_{2\times S_i} = \begin{bmatrix} \theta^{(i)}_{11} & \theta^{(i)}_{12} & \cdots & \theta^{(i)}_{1S_i} \\ \theta^{(i)}_{21} & \theta^{(i)}_{22} & \cdots & \theta^{(i)}_{2S_i} \end{bmatrix} \in \mathbb{R}_+^{2\times S_i}$ |
| 09 $\qquad\quad \theta_0 = u_2(\theta_H - \theta_L) + \theta_L$ | |
| 10 $\qquad\quad t_2 = t_2 + k$ | $\aleph_{1\times\gamma} = [n_1 \; n_2 \; \cdots \; n_\gamma] \in \mathbb{N}_*^\gamma$ |
| 11 $\qquad\quad P_{1\times 2}^{(t_2)} = [r_0 \cos(\theta_0) \; r_0 \sin(\theta_0)]$ | |
| 12 $\qquad$ **end** | **Outputs** |
| 13 $\quad$ **end** | 1 – CSA Matrix 6 – Limits of a Specific |
| 14 $\quad t_1 = t_1 + S_i$ | 2 – Zeros Vector  Sector [each sector is |
| 15 **end** | 3 – Radii Matrix   within 2 radii and 2 |
| | 4 – Angles Matrix  angles]. |
| | 5 – Nodes Vector |

## V. CONCLUSION

In this contribution we derived *exact* stochastic expressions for node scattering and PL density, unbiased and more generic and than those currently employed. We then used the results and showed CSA, a more flexible spatial algorithm for non-homogenous node emulation. Last, simulation was used to reaffirm the validity of theoretical analysis.